\documentclass[runningheads,a4paper]{llncs}

\usepackage{amssymb}
\usepackage{exscale}
\usepackage[centertags]{amsmath}
\usepackage{amssymb}
\usepackage{makeidx}         
\usepackage{graphicx}                                  
\usepackage{multicol}        
\usepackage[bottom]{footmisc}
\usepackage{amsfonts}
\usepackage{amsmath}
\usepackage{dcolumn}
\usepackage{multirow}
\usepackage{subfig}
\usepackage{algorithmicx}
\usepackage{algpseudocode}
\usepackage{algorithm}
\usepackage{color,,stmaryrd}

\usepackage{url}
\newcommand*{\defeq}{\mathrel{\vcenter{\baselineskip0.5ex \lineskiplimit0pt
      \hbox{\footnotesize.}\hbox{\footnotesize.}}}%
  =}

\newcommand{\given}{\, | \,}

\newcommand{\incomp}{\sim_U}
\newcommand{\indiff}{\sim_E}
\newcommand{\dX}{\mathbb{X}}
\newcommand{\dY}{\mathbb{Y}}

\newcommand{\vek}[1]{\boldsymbol{#1}}
\newcommand{\on}[1]{\operatorname{#1}}
\newcommand{\sign}{\on{sign}}

\begin{document}

\mainmatter  % start of an individual contribution

\title{Weighting by Tying: A New Approach to Weighted Rank Correlation}

\titlerunning{Weighted Rank Correlation}
\author{Sascha Henzgen\inst{1} and Eyke H\"ullermeier\inst{2,3}}
%\thanks{...}
%

\authorrunning{Sascha Henzgen and Eyke H\"ullermeier}

\institute{pmOne Group\\
Paderborn, Germany
\and
Institute of Informatics, University of Munich (LMU)\\
\and
Munich Center for Machine Learning (MCML)\\
Munich, Germany
}

\toctitle{Weighted Rank Correlation}
\tocauthor{Authors' Instructions}
\maketitle

\begin{abstract}
Measures of rank correlation are commonly used in statistics to capture the degree of concordance between two orderings of the same set of items. Standard measures like Kendall's tau and Spearman's rho coefficient put equal emphasis on each position of a ranking. Yet, motivated by applications in which some of the positions (typically those on the top) are more important than others, a few weighted variants of these measures have been proposed. Most of these generalizations fail to meet desirable formal properties, however. Besides, they are often quite inflexible in the sense of committing to a fixed weighing scheme. In this paper, we propose a weighted rank correlation measure on the basis of fuzzy order relations. Our measure, called \emph{scaled gamma}, is related to Goodman and Kruskal's gamma rank correlation. It is parametrized by a fuzzy equivalence relation on the rank positions, which in turn is specified conveniently by a so-called scaling function. This approach combines soundness with flexibility: it has a sound formal foundation and allows for weighing rank positions in a flexible way. 
%The usefulness of our class of weighted rank correlation measures is shown by means of experimental studies using both synthetic and real-world ranking data. 
\end{abstract}

\section{Introduction}

Rank correlation measures  such as Kendall's tau \cite{kend_rc} and Spearman's rho \cite{spea_tp04}, which have originally been developed in non-parametric statistics, are used extensively in various fields of application, ranging from bioinformatics \cite{mpub112} to information retrieval \cite{Yilmaz08}.  In contrast to numerical correlation measures such as Pearson correlation, rank correlation measures are only based on the ordering of the observed values of a variable. Thus, measures of this kind are not limited to numerical variables but can also be applied to non-numerical variables with an ordered domain (i.e., measured on an ordinal scale) and, of course, to rankings (permutations) directly.

In many applications, such as Internet search engines, one is not equally interested in all parts of a ranking. Instead, the top positions of a ranking (e.g., the first 10 or 50 web sites listed) are typically considered more important than the middle part and the bottom. Standard rank correlation measures, however, put equal emphasis on all positions. Therefore, they cannot distinguish disagreements  in different parts of a ranking. This is why \emph{weighted} variants have been proposed for some correlation measures, as well as alternative measures specifically focusing on the top of a ranking \cite{CostaSoares2005,Fagin2003,Kaye1973,Maturi2008,Yilmaz08}. Most of these generalizations fail to meet desirable formal properties, however. Besides, they are often quite inflexible in the sense of committing to a fixed weighing scheme.

In this paper, we develop a general framework for designing weighted rank correlation measures based on the notion of \emph{fuzzy order relation}, and use this framework to generalize Goodman and Kruskal's gamma coefficient \cite{Goodman79}. Our approach has a sound formal foundation and allows for weighing rank positions in a flexible way. In particular, it is not limited to monotone weighing schemes that emphasize the top in comparison to the rest of a ranking. The key ingredients of our approach, to be detailed further below, are as follows:
\begin{itemize}
\item \emph{Fuzzy order relations} \cite{bode_sf08} are generalizations of the conventional order relations on the reals or the integer numbers: $\mathtt{SMALLER}$, $\mathtt{EQUAL}$ and $\mathtt{GREATER}$. They enable a smooth transition between these predicates and allow for expressing, for instance, that a number $x$ is smaller than $y$ \emph{to a certain degree}, while to some degree these numbers are also considered as being equal. Here, the $\mathtt{EQUAL}$ relation is understood as a kind of similarity relation that seeks to model the ``perceived equality'' (instead of the strict mathematical equality).

\item \emph{Scaling functions} for modeling fuzzy equivalence relations \cite{klaw_fs}. For each element $x$ of a linearly ordered domain $X$, a scaling function $s(\cdot)$ essentially expresses the degree $s(x)$ to which $x$ can be (or should be) distinguished from its neighboring values. A measure of distance (or, equivalently, of similarity) on $X$ can then be derived via accumulation of local degrees of distinguishability.

\item \emph{Fuzzy rank correlation} \cite{bode_rr08,mpub241} generalizes conventional rank correlation on the basis of fuzzy order relations, thereby combining properties of standard rank correlation (such as Kendall's tau) and numerical correlation measures (such as Pearson correlation). 
Roughly, the idea is to penalize the inversion of two items (later on called a \emph{discordance}) depending on how dissimilar the corresponding rank positions are: the more similar (less distinguishable) the positions are according to the $\mathtt{EQUAL}$ relation, the smaller the influence of the inversion on the rank correlation. 

\end{itemize}
The rest of the paper is organized as follows. In the next two sections, we briefly recall the basics of fuzzy order relations and fuzzy rank correlation, respectively. Our weighted rank correlation measure, called \emph{scaled gamma}, is then introduced in Section 5, and related work is reviewed in Section 6.  A small experimental study is presented in Section 7, prior to concluding the paper in Section 8.

\section{Rank Correlation}

Consider $N \geq 2$ paired observations $\{ (x_i,y_i) \}_{i=1}^N   \subset   \dX \times \dY$
of two variables $X$ and $Y$, where $\dX$ and $\dY$ are two linearly ordered domains; we denote 
\[
\vek{x} = (x_1, x_2, \ldots, x_N) , \enspace 
\vek{y} = (y_1, y_2, \ldots, y_N) \enspace .
\]
In particular, the values $x_i$ (and $y_i$) can be real numbers ($\dX = \mathbb{R}$) or rank positions ($\dX =  [N] \defeq \{ 1, 2, \ldots , N \}$). For example, $\vek{x} = (3,1,4,2)$ denotes a ranking of four items, in which the first item is on position 3, the second on position 1, the third on position 4 and the fourth on position 2.

The goal of a (rank) correlation measure is to capture the dependence between the two variables in terms of their tendency to increase and decrease (their position) in the same or the opposite direction. If an increase in $X$ tends to come along with an increase in $Y$, then the (rank) correlation is positive. The other way around, the correlation is negative if  an increase in $X$ tends to come along with a decrease in $Y$. If there is no dependency of either kind, the correlation is (close to) 0. 

\subsection{Concordance and Discordance}

Many rank correlation measures are defined in terms of the number $C$ of \emph{concordant}, the number $D$ of \emph{discordant}, and the number $T$ of \emph{tied} data points. Let $\mathcal{P} = \{ (i,j) \, \vert \, 1 \leq i < j \leq N \}$ denote the set of ordered index pairs. We call a pair $(i,j) \in \mathcal{P}$  concordant, discordant or tied depending on whether 
$(x_i-x_j)(y_i-y_j)$ is positive, negative or 0, respectively. Thus, let us define the following $N \times N$ relations:
% $\mathcal{C}$, $\mathcal{D}$ and $\mathcal{T}$ as follows:
\begin{align}
\label{eq:relc}
\mathcal{C}(i,j) & = \left\{ \begin{array}{cl}
1 & \text{ if } \sign(\pi_x(i)-\pi_x(j))\sign(\pi_y(i)-\pi_y(j)) = 1 \\
0 & \text{ otherwise} 
\end{array} \right. \\[3mm]
\label{eq:reld}
\mathcal{D}(i,j) & = \left\{ \begin{array}{cl}
1 & \text{ if } \sign(\pi_x(i)-\pi_x(j))\sign(\pi_y(i)-\pi_y(j)) = -1 \\
0 & \text{ otherwise} 
\end{array} \right. \\[3mm]
\label{eq:reltx}
\mathcal{T}_x(i,j) & = \left\{ \begin{array}{cl}
1 & \text{ if } \sign(\pi_x(i)-\pi_x(j))=0, \, \sign(\pi_y(i)-\pi_y(j)) \neq 0 \\
0 & \text{ otherwise} 
\end{array} \right.\\[3mm]
\label{eq:relty}
\mathcal{T}_y(i,j) & = \left\{ \begin{array}{cl}
1 & \text{ if } \sign(\pi_x(i)-\pi_x(j))\neq 0, \, \sign(\pi_y(i)-\pi_y(j))= 0 \\
0 & \text{ otherwise} 
\end{array} \right.\\[3mm]
\label{eq:reltxy}
\mathcal{T}_{x,y}(i,j) & = \left\{ \begin{array}{cl}
1 & \text{ if } \sign(\pi_x(i)-\pi_x(j)) = 0, \, \sign(\pi_y(i)-\pi_y(j))= 0 \\
0 & \text{ otherwise} 
\end{array} \right.\\[3mm]
\label{eq:relt}
\mathcal{T}(i,j) & = \left\{ \begin{array}{cl}
1 & \text{ if } \sign(\pi_x(i)-\pi_x(j)) \sign(\pi_y(i)-\pi_y(j))= 0 \\
0 & \text{ otherwise} 
\end{array} \right.
\end{align}
Obviously, $\mathcal{T} = \mathcal{T}_x \cup \mathcal{T}_y \cup \mathcal{T}_{x,y}$.

The number of concordant, discordant and tied pairs $(i,j) \in \mathcal{P}$ are then obtained by summing the entries in the corresponding relations:
\begin{align*}
C & = \sum_{(i,j) \in \mathcal{P}} \mathcal{C}(i,j) = \frac{1}{2} \sum_{i \in [N]} \sum_{j \in [N]} \mathcal{C}(i,j) \\
D & =\sum_{(i,j) \in \mathcal{P}} \mathcal{D}(i,j) = \frac{1}{2} \sum_{i \in [N]} \sum_{j \in [N]} \mathcal{D}(i,j) \\
T & =\sum_{(i,j) \in \mathcal{P}} \mathcal{T}(i,j) = \frac{1}{2} \sum_{i \in [N]} \sum_{j \in [N]} \mathcal{T}(i,j) - \frac{N}{2}
\end{align*}
Note that 
\begin{equation}\label{eq:part}
\mathcal{C}(i,j) + \mathcal{D}(i,j) + \mathcal{T}(i,j) = 1 
\end{equation}
for all $(i,j) \in \mathcal{P}$, and 
\begin{equation}\label{eq:part2}
C + D + T =  | \mathcal{P} | = \frac{N(N-1)}{2} \enspace .
\end{equation}

Well-known examples of rank correlation measures that can be expressed in terms of the above quantities include Kendall's tau \cite{kend_rc}
\begin{equation}
\tau = \frac{C-D}{N(N-1)/2}
\label{eq:kendall}
\end{equation}
and Goodman and Kruskal's gamma coefficient \cite{Goodman79}
\begin{equation}
\gamma = \frac{C-D}{C+D}\enspace .
\label{eq:gamma}
\end{equation}
As will be detailed in the following sections, our basic strategy for generalizing rank correlation measures such as $\gamma$ is to ``fuzzify'' the concepts of concordance and discordance. Thanks to the use of fuzzy order relations, we will be able to express that a pair $(i,j)$ is concordant or discordant to a certain degree (between $0$ and $1$). Measures like (\ref{eq:gamma}) can then be generalized in a straightforward way, namely by accumulating the degrees of concordance and discordance, respectively, and putting them in relation to each other.

\subsection{Rank Correlation and Distance}

Formally, there is a close connection between rank correlation and distance measures on rankings. Indeed, many rank correlation measures are in fact normalized versions of corresponding distance measures. For example, Spearman's rho is an affine transformation of the sum of squared rank distances to the interval $[-1,+1]$, and Kendall's tau is a similar transformation of the Kendall distance, namely the sum of rank inversions. In general, the relationship between a correlation $\on{corr}$ and a distance $d$ is expressed as follows:
\begin{equation}\label{eq:cd}
\on{corr}(\pi_x , \pi_y) = 1 - \frac{2 d(\pi_x, \pi_y)}{M} \, ,
\end{equation}
where $M$ is a maximal distance. Desirable formal properties of a measure are more naturally given in terms of distance. For example, the following axioms were proposed by Kemmey and Snell \cite{edmo_an02}:
\begin{itemize}
	\item[A1] The distance $d$ is a metric, i.e., the following properties hold for all $\pi_x, \pi_y, \pi_z$:
		\begin{itemize}
			\item[--] $d(\pi_x,\pi_x) \geq 0$  
			\item[--] $(\pi_x = \pi_y)$ implies  $d(\pi_x,\pi_y) = 0$ 			
			\item[--] $d(\pi_x,\pi_y) = 0$ implies $\pi_x = \pi_y$ 
			\item[--] $d(\pi_x,\pi_y) = d(\pi_x,\pi_y)$  
			\item[--] $d(\pi_x,\pi_z) \leq d(\pi_x,\pi_y) + d(\pi_y,\pi_z)$.
			%and the equality holds if and only if the ranking $\pi_y$ is between $\pi_1$ and $\pi_3$.
		\end{itemize}
		\item[A2] Re-labeling: If $\pi_x'$ results from $\pi_x$ by a permutation of the items, and $\pi_y'$ results from $\pi_y$ by the same permutation, then $d(\pi_x',\pi_y') = d(\pi_x,\pi_y)$ .
		\item[A3] If two rankings $\pi_x$ and $\pi_y$ agree on all except a subset $S$ of items, and these items form a segment in both rankings (i.e., they occupy a sequence of consecutive positions), then $d(\pi_x,\pi_y)$ equals the distance on the subset $S$ (i.e., the rankings $\pi_x$ and $\pi_y$ projected to $S$).
		
		\item[A4] The minimum positive distance is 1.
\end{itemize}
%For Axiom 1.3 it has to be clarified, when a ranking $\pi_2$ is between $\pi_1$ and $\pi_3$. Kemeny and Snell defined $\pi_2$ to be between $\pi_1$ and $\pi_3$ if for every pair $\{i,j\}$, $\pi_2$ agrees with either $\pi_1$ or $\pi_3$ or, if $\pi_1$ and $\pi_3$ are not agreeing, $i$ and $j$ are tied in $\pi_2$.

\section{Fuzzy Relations}

\subsection{Fuzzy Equivalence}

The notion of a fuzzy relation generalizes the standard notion of a mathematical relation by allowing one to express ``degrees of relatedness''.  Formally, a (binary) fuzzy relation on a set $\mathbb{X}$ is characterized by a membership function 
$$
\mathcal{E}: \mathbb{X} \times \mathbb{X} \longrightarrow \left[0,1\right] \, .
$$ 
For all $x , y \in \mathbb{X}$, $\mathcal{E}(x,y)$ is the degree to which $x$ is related to $y$. 

Recall that a conventional equivalence relation on a set $\mathbb{X}$ is a binary relation that is reflexive, symmetric and transitive. For the case of a fuzzy relation $\mathcal{E}$, these properties are generalized as follows:	
	\begin{itemize}
		\item reflexivity: $\mathcal{E}(x,x) = 1$ for all $x \in \mathbb{X}$
		\item symmetry: $\mathcal{E}(x,y) = \mathcal{E}(y,x)$ for all $x,y  \in \mathbb{X}$
		\item $\top$-transitivity: $\top(\mathcal{E}(x,y), \mathcal{E}(y,z)) \leq \mathcal{E}(x,z)$ for all $x,y,z \in \mathbb{X}$
	\end{itemize}
A fuzzy relation $\mathcal{E}$ having these properties is called a fuzzy equivalence relation \cite{bode_rr08}. 
While the generalizations of reflexivity and symmetry are rather straightforward, the generalization of transitivity involves a triangular norm (t-norm) $\top$, which plays the role of a generalized logical conjunction \cite{klem_tn}. Formally, a function $\top: \, [0,1]^2 \longrightarrow [0,1]$ is a t-norm if it is associative, commutative, monotone increasing in both arguments, and satisfies the boundary conditions $\top(a,0)=0$ and $\top(a,1)=a$ for all $a \in [0,1]$. Examples of commonly used t-norms include the minimum $\top(a,b)=\min(a,b)$ and the product $\top(a,b)=ab$. To emphasize the role of the t-norm, a relation $\mathcal{E}$ satisfying the above properties is also called a $\top$-equivalence.

\subsection{Fuzzy Ordering}

The notion of an order relation $\leq $ is similar to that of an equivalence relation, with the important difference that the former is antisymmetric while the latter is symmetric. A common way to formalize antisymmetry is as follows: $a \leq b$ and $b \leq a$ implies $a=b$. Note that this definition already involves an equivalence relation, namely the equality $=$ of two elements. Thus, as suggested by Bodenhofer \cite{Bodenhofer00}, a fuzzy order relation can be defined on the basis of a fuzzy equivalence relation. Formally, a fuzzy relation $\mathcal{L}: \mathbb{X} \times \mathbb{X} \longrightarrow \left[0,1\right]$ is called a \emph{fuzzy ordering} with respect to a t-norm $\top$ and a $\top$-equivalence $\mathcal{E}$, for brevity $\top$-$\mathcal{E}$-$ordering$, if it satisfies the following properties for all $x,y,z \in \mathbb{X}$:
\begin{itemize}
	\item $\mathcal{E}$-reflexivity: $\mathcal{E}(x,y) \leq \mathcal{L}(x,y)$ 
	\item $\top$-$\mathcal{E}$-antisymmetry: $\top(\mathcal{L}(x,y),\mathcal{L}(y,x))\leq \mathcal{E}(x,y)$ 
	\item $\top$-transitivity: $\top(\mathcal{L}(x,y), \mathcal{L}(y,z)) \leq \mathcal{L}(x,z)$ 
	\end{itemize}
Furthermore a $\top$-$\mathcal{E}$-ordering $\mathcal{L}$ is called \textit{strongly complete} if 
$$
\max \big(\mathcal{L}(x,y),\mathcal{L}(y,x) \big)= 1
$$
for all $x,y \in \mathbb{X}$. This is expressing that, for each pair of elements $x$ and $y$, either $x \leq y$ or $y \leq x$ should be fully true. 

A fuzzy relation $\mathcal{L}$ as defined above can be seen as a generalization of the conventional ``smaller or equal'' on the real or the integer numbers. What is often needed, too, is a ``stricly smaller'' relation $<$. In agreement with the previous formalizations, a relation of that kind can be defined as follows: A binary fuzzy relation $\mathcal{R}$ is called a \emph{strict fuzzy ordering} with respect to a $\top$-norm and a $\top$-equivalence $\mathcal{E}$, or strict $\top$-$\mathcal{E}$-$ordering$ for short, if it has the following properties for all $x,x',y,y',z \in \mathbb{X}$~\cite{bode_rr08}:
\begin{itemize}
\item irreflexivity: $\mathcal{R}(x,x) = 0$ 
\item $\top$-transitivity: $\top(\mathcal{R}(x,y), \mathcal{R}(y,z)) \leq \mathcal{R}(x,z)$ 
\item $\mathcal{E}$-extensionality: $\top(\mathcal{E}(x,x'),\mathcal{E}(y,y'), \mathcal{R}(x,y)) \leq \mathcal{R}(x',y')$ 
\end{itemize}

\subsection{Practical Construction}
\label{sec:practical}

The above definitions provide generalizations $\mathcal{E}$, $\mathcal{L}$ and $\mathcal{R}$ of the standard relations $=$, $\leq$ and $<$, respectively, that exhibit reasonable properties and, moreover, are coherent with each other. Practically, one may start by choosing an equivalence relation $\mathcal{E}$ and a compatible t-norm $\top$, and then derive $\mathcal{L}$ and $\mathcal{R}$ from the corresponding $\top$-equivalence.  

More specifically, suppose the set $\mathbb{X}$ to be a linearly ordered domain, that is, to be equipped with a standard (non-fuzzy) order relation $\leq$. Then, given a  $\top$-equivalence $\mathcal{E}$ on $\mathbb{X}$, the following relation is a coherent fuzzy order relation, namely a strongly complete $\top$-$\mathcal{E}$-ordering:
$$
   \mathcal{L}(x,y)= \left\{ \begin{array}{cl}
            1 &\text{if } x \leq y \\
            \mathcal{E}(x,y) &\text{otherwise } 
            \end{array} \right.
$$
Moreover, a strict fuzzy ordering $\mathcal{R}$ can be obtained from $\mathcal{L}$ by
\begin{equation}\label{eq:r}
\mathcal{R}(x,y) = 1 - \mathcal{L}(y,x) 
\end{equation}
The relations thus defined have a number of convenient properties. In particular, $\min ( \mathcal{R}(x,y), \mathcal{R}(y,x) ) = 0$ and
\begin{equation}\label{eq:partition}
\mathcal{R}(x,y)+ \mathcal{E}(x,y) + \mathcal{R}(y,x)= 1
\end{equation}
for all $x, y \in \mathbb{X}$. These properties can be interpreted as follows. For each pair of elements $x$ and $y$, the unit mass splits into two parts: a degree $a = \mathcal{E}(x,y)$ to which $x$ and $y$ are equal, and a degree $1-a$ to which either $x$ is smaller than $y$ or $y$ is smaller than $x$.

\section{Fuzzy Relations on Rank Data}

Since we are interested in generalizing rank correlation measures, the underlying domain $\mathbb{X}$ is given by a set of rank positions  $[N]$ (equipped with the standard $<$ relation) in our case. As mentioned before, this domain could be equipped with fuzzy relations $\mathcal{E}$, $\mathcal{L}$ and $\mathcal{R}$ by defining $\mathcal{E}$ first and deriving $\mathcal{L}$ and $\mathcal{R}$ afterward. Note, however, that the number of degrees of freedom in the specification of $\mathcal{E}$ is of the order $O(N^2)$, despite the constraints this relation has to meet. 

\subsection{Scaling Functions on Rank Positions}

In order to define fuzzy relations even more conveniently, while emphasizing the idea of weighing the importance of rank positions at the same time, we leverage the concept of a \emph{scaling function} as proposed by Klawonn \cite{klaw_fs}. Roughly speaking, a scaling function $w:\, \mathbb{X} \longrightarrow \mathbb{R}_+$ specifies the dissimilarity of an element $x$ from its direct neighbor elements, and the dissimilarity between any two elements $x$ and $y$ is then obtained via integration of the local dissimilarities along the chain from $x$ to $y$. In our case, a scaling function can be defined as a mapping $w: \, [N-1] \longrightarrow [0,1]$ or, equivalently, as a vector 
\begin{equation}\label{eq:s}
\vec{w} = \Big( w(1), w(2), \ldots , w(N-1) \Big)  \in [0,1]^{N-1} \enspace .
\end{equation}
Here, $w(n)$ can be interpreted as the degree to which the rank positions $n$ and $n+1$ are distinguished from each other; correspondingly, $1-w(n)$ can be seen as the degree to which these two positions are considered to be equal. From the local degrees of distinguishability, a global distance function is derived on $\mathbb{X}$ by defining
\begin{equation}\label{eq:sum}
d(x,y) = \min \left(  1, \sum_{i=\min(x,y)}^{\max(x,y)-1} w(i)   \right) \enspace .
\end{equation}
Put in words, the distance between $x$ and $y$ is the sum of the degrees of distinguishability between them, thresholded at the maximal distance of 1. In principle, accumulations of the degrees of distinguishability other than the sum are of course conceivable. For example, the maximum could be used as well: 
\begin{equation}
d(x,y) = \max \Big\{ w(i) \given i \in \{ \min(x,y), \ldots , \max(x,y)-1 \}  \Big\} \enspace .
\label{eq:maxAgg}
\end{equation}
In general, $d(x,y)$ is supposed to define a pseudo-metric on $\mathbb{X}$. Under this condition, it can be shown that the fuzzy relation $\mathcal{E}$ defined as
$$
\mathcal{E}(x,y) = 1- d(x,y) 
$$
for all $x,y \in \mathbb{X}$ is a $\top_L$-equivalence, where $\top_L$ is the  {\L}ukasiewicz $t$-norm $\top_L(a,b) = \max \{ 0, a+b-1 \}$ \cite{Bodenhofer03}. Relations $\mathcal{L}$ and $\mathcal{R}$ can then be derived from $\mathcal{E}$ as described in Section \ref{sec:practical}. In particular, we obtain
$$
\mathcal{R}(x,y) = \left\{ \begin{array}{cl}
d(x,y) & \text{ if } x < y \\
0 & \text{ otherwise }  
\end{array} \right.
$$
According to our discussion so far, the only remaining degree of freedom is the scaling function $s$. Obviously, this function can also be interpreted as a \emph{weighing function}: the more distinguishable a position $n$ from its neighbor positions, i.e., the larger $w(n-1)$ and $w(n)$, the higher the importance of that position.

\begin{figure}
	\centering
		\includegraphics[width=0.70\textwidth]{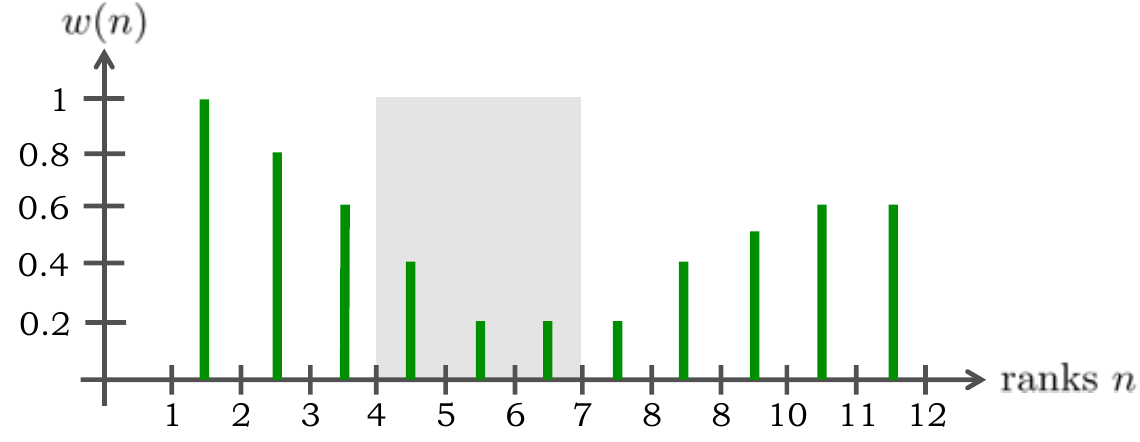}
		\caption{Example of a scaling function.}
			\label{fig:scaling}
\end{figure}

An example of a scaling function for $N=12$ is shown in Figure \ref{fig:scaling}. This function puts more emphasis on the top and the bottom ranks and less on the middle part. According to (\ref{eq:sum}), the distinguishability between the positions $4$ and $7$ is $d(4,7)=0.4+0.2+0.2=0.8$ (sum of the weights $w(i)$ in the shaded region). Thus, $4$ is strictly smaller than $7$ to the degree of $\mathcal{R}(4,7)=0.8$, while both positions are considered equal to the degree $\mathcal{E}(4,7) = 0.2$.

Note that, with $w(i)= \llbracket i < k \rrbracket$, we also cover the top-$k$ scenario as a special case. Here, the standard $<$ relation is recovered for all elements on the first $k$ positions, whereas the remaining positions are considered as fully equivalent, i.e., these elements form an equivalence class in the standard sense.

\section{Weighted Rank Correlation}

Our approach to generalizing rank correlation measures is based on the ``fuzzificiation'' of the relations (\ref{eq:relc}--\ref{eq:relt}) and, correspondingly, the number of concordant, discordant and tied item pairs. The tools that are needed to do so have already been introduced in the previous sections. In particular, suppose a fuzzy equivalence relation $\mathcal{E}$ and a ``strictly smaller'' relation $\mathcal{R}$ to be derived from a scaling function $w$ on $\mathbb{X}$, based on the procedure outlined above. For notational convenience, we assume the same scaling function (and hence the same relations) to be used on both domains $\mathbb{X}$ and $\mathbb{Y}$. In principle, however, different functions $w_X$ and $w_Y$ (and hence relations $\mathcal{E}_X$, $\mathcal{R}_X$ and $\mathcal{E}_Y$, $\mathcal{R}_Y$) could be used. 

Now, according to (\ref{eq:relc}), a pair $(i,j) \in \mathcal{P}$ is concordant if both $x_i$ is (strictly) smaller than $x_j$ and $y_i$ is smaller than $y_j$, or if $x_j$ is smaller than $x_i$ and $y_j$ is smaller than $y_i$. Using our fuzzy relation $\mathcal{R}$ and a t-norm $\top$ as a generalized conjunction, this can be expressed as follows:
\begin{equation}
\tilde{\mathcal{C}}(i,j) = \top \big( \mathcal{R}(x_i , x_j) , \mathcal{R}(y_i , y_j) \big) +  \top \big( \mathcal{R}(x_j, x_i) , \mathcal{R}(y_j , y_i) \big)
\label{eq:fuzzyC}
\end{equation}
The discordance relation can be expressed analogously:
\begin{equation}
\tilde{\mathcal{D}}(i,j) = \top \big( \mathcal{R}(x_i , x_j) , \mathcal{R}(y_j , y_i) \big) +  \top \big( \mathcal{R}(x_j, x_i) , \mathcal{R}(y_i, y_j) \big)
\label{eq:fuzzyD}
\end{equation}
Finally, the degree to which $(i,j)$ is tied is given by
$$
\tilde{\mathcal{T}}(i,j) = \bot \big( \mathcal{E}(x_i , x_j) , \mathcal{E}(y_i , y_j) \big) \enspace ,
$$
where $\bot$ is the t-conorm associated with $\top$ (i.e., $\bot(u,v)= 1-\top(1-u,1-v)$), serving as a generalized logical disjunction. Generalizing (\ref{eq:part}), the three degrees sum up to 1, i.e., 
\begin{equation}\label{eq:fpart}
\tilde{\mathcal{C}}(i,j) + \tilde{\mathcal{D}}(i,j)  + \tilde{\mathcal{T}}(i,j) \equiv 1 \enspace ,
\end{equation}
and either $\tilde{\mathcal{C}}(i,j) = 0$ or $\tilde{\mathcal{D}}(i,j) = 0$. In other words, a pair $(i,j)$ that has originally been concordant (discordant) will remain concordant (discordant), at least to some extent. However, since $\mathcal{E}$ may introduce a certain indistinguishability between the positions $x_i$ and $x_j$ or the positions $y_i$ and $y_j$, the pair could also be considered as a partial tie.

Given the above fuzzy relations, the number of concordant, discordant and tied data points can be obtained as before, namely by summing over all ordered pairs $(i,j) \in \mathcal{P}$:
$$
\tilde{C} = \sum_{(i,j) \in \mathcal{P}} \tilde{\mathcal{C}}(i,j) \enspace , \quad
\tilde{D} = \sum_{(i,j) \in \mathcal{P}} \tilde{\mathcal{D}}(i,j) \enspace , \quad
\tilde{T} = \sum_{(i,j) \in \mathcal{P}} \tilde{\mathcal{T}}(i,j) \enspace .
$$
According to (\ref{eq:fpart}), 
$$
\tilde{C} + \tilde{D} + \tilde{T}  = | \mathcal{P} | = \frac{N(N-1)}{2} \enspace ,
$$
which generalizes (\ref{eq:part2}). Using these quantities, rank correlation measures expressed in terms of the number of concordant and discordant pairs can be generalized in a straightforward way. In particular, a generalization of the gamma coefficient (\ref{eq:gamma}) is obtained as
\begin{equation}\label{eq:fgamma}
\tilde{\gamma}  = \frac{\tilde{C} - \tilde{D}}{\tilde{C} + \tilde{D}} \enspace .
\end{equation}
It is worth mentioning that the weighted rank correlation measure thus defined exhibits a number of desirable formal properties, which it essentially inherits from the general fuzzy extension of the gamma coefficient; we refer to  \cite{mpub241}, in which these properties are analyzed in detail.

\section{Dealing with Ties}

The basic principle of our approach to positional weighing of rank correlation measures is the idea of discounting concordances and discordances of pairs of items by turning them into \emph{partial ties}: If the positions occupied by the items are considered as (partly) equivalent according to the underlying equivalence relation on ranks, a concordance (discordance) is turned into a partial concordance (discordance) and a partial tie. The way in which ties are dealt with by the subsequently applied rank correlation measure is hence very critical:
\begin{itemize}
\item To achieve the desired weighting effect, the effect of ties on the overall correlation should be less strong than the effect of concordances or discordances.
\item In this regard, it is important to realize that ties are indeed treated differently by different measures, which is also due to different interpretations of ties. 
\item This issue becomes especially critical if we seek to apply our approach to rankings that may already contain ties (which we excluded so far). In this case, there are two types of ties: ``real ties'' contained in the original rankings and ``artificial ties'' resulting from our transformation. These two types should arguably be distinguished. 
\end{itemize}

\subsection{Notation}

As before, we consider data in the form of objects $\vek{x} = (x_1, x_2, \ldots, x_N) \in \dX = [N] \defeq \{ 1, 2, \ldots , N \}$. Now, however, we do not necessarily assume that all entries $x_i$ are distinct. Instead, if $x_i = x_j$ for $i \neq j$, we consider the two items to be tied. For example, denoting the items to be ranked by $A, B, C$, etc., $\vec{x} = (2,1,3,2,1)$ represents the ranking $\{B, E\} \succ \{A,D\} \succ C$ for items $\{A,B,C,D,E\}$, which we can also write as $BE|AD|C$.

As before, the preference relation $\succ$ on items is derived from the standard $<$ relation on $\dX$. The set of items with the same rank form a tie group or equivalence class. We denote by $[n] = \{ i \given x_i = n \}$ the set of items (or rather their indices) put on position $n$. The preference $\succ$ thus defined is not only irreflexive, asymmetric, and transitive, but also satisfies transitivity of incomparability $\sim$, where $A \sim B$ if neither $A \succ B$ nor $B \succ A$: If $A \sim B$ and $B \sim C$, then $A \sim C$. In the literature, such relations are called strict weak order, or sometimes ``bucket orders'' (with the idea that all items within an equivalence class are put in a bucket).

%Extending $\dX$ by an additional element $\bot$, we can also express that an item is not ranked at all, i.e., we can model incomplete rankings. For example, $\vek{x} = (2,1,3,\bot, 2)$ represents the ranking $B \succ \{A,E\} \succ C$, in which item $D$ does not occur.   Extending the order on $\dX$ by assuming that neither $\bot < n$ nor $\bot > n$ for any $n \in [N]$, the induced preference $\succ$ on items is 

\subsection{Interpretation of Ties}

There are two main interpretations of ties:
\begin{itemize}
\item Indifference ($\indiff$): First, two items can be tied because they are indeed considered as equal ($\sim$ is interpreted as $\indiff$). For example, a decision maker might be indifferent between two choice alternatives, because she likes them both the same. 
\item Incomparability ($\incomp$): Second, two items could be tied in a ranking because their true relationship is unknown, i.e., they are considered as incomparable due to a lack of information ($\sim$ is interpreted as $\incomp$). 
\end{itemize}
It is important to notice that, according to the first interpretation, a tie $A \sim B$ directly refers to properties of the underlying items $A$ and $B$ (it has an \emph{ontic} meaning), whereas the second interpretation is of \emph{epistemic} nature and represents \emph{knowledge} about the relation between the items: $A \sim B$ does not mean that $A$ and $B$ are equally preferred. Instead, given the information at hand, we simply cannot compare them, i.e., we cannot determine the ``true'' relationship between $A$ and $B$. Also note that, regarding this ``ground truth'', different assumptions could be made: We may assume that the ground truth is a total order, i.e., either $A \succ B$ or $B \succ A$, we only do not know (yet). Or, we may even allow that the ground truth is an indifference between $A$ and $B$. In this case, the two different interpretations of ties would co-exist: $A \incomp B$ means that $A \succ B$ or $B \succ A$ or $A \indiff B$. Obviously, this co-existence will lead to additional complications. Therefore, whenever the interpretation of ties in terms of incomparability is adopted, we assume that the ground truth is a total order (ranking without indifference).

\subsection{Treatment of Ties}

Naturally, the different interpretations of ties call for different ways of treating them in the context of rank correlation. For example, $A \indiff B$ in two rankings suggests a strong coherence, because both rankings agree on the equality of $A$ and $B$, and this coherence should be rewarded by a correlation measure. In the case of incomparability, the situation is slightly different: $A \incomp B$ in two observed rankings could indicate both a concordance and a discordance in the underlying ground truth rankings. In this situation, it might be natural to express \emph{uncertainty} about the actual correlation.

More generally, the interpretation of ties in terms of indifference calls for an extended scoring scheme, which, in addition to (scores for) concordance and discordance also considers comparisons involving ties:

\begin{center}
\begin{tabular}{c|ccc}
  & \quad$\pi_y(i) < \pi_y(j)$\quad & \quad$\pi_y(i) = \pi_y(j)$\quad & \quad$\pi_y(i) > \pi_y(j)$\quad \\
  \hline
$\pi_x(i) < \pi_x(j)$ & $s_{1,1}$ & $s_{1,2}$ & $s_{1,3}$ \\  
$\pi_x(i) = \pi_x(j)$  & $s_{2,1}$ & $s_{2,2}$ & $s_{2,3}$ \\  
$\pi_x(i) > \pi_x(j)$  & $s_{3,1}$ & $s_{3,2}$ & $s_{3,3}$ \\
\end{tabular}
\end{center}
Naturally, the highest scores (rewards) should be found on the diagonal, whereas $s_{1,3}$ and $s_{3,1}$ should be lowest; the values $s_{1,2}, s_{2,1}, s_{2,3}, s_{3,2}$ should be in-between, because they can be seen as half a concordance and half a discordance. 

In the case of incomparability, it is natural to compare two rankings $\pi$ and $\pi'$ by considering the set of possible values of correlation between the underlying (but unknown) ground truth rankings, that is, the set of extensions $E(\pi)$ and $E(\pi')$. For a correlation measure $\on{corr}$, which only needs to be defined on complete rankings, this set if given by
$$
\Big\{  \on{corr}(\tau , \tau') \given \tau \in E(\pi) , \, \tau' \in E(\pi') \Big\} \, .
$$
To obtain a single value, the values in this set could be aggregated in one way or the other, for example in terms of the minimum (reflecting a pessimistic attitude), the maximum (reflecting an optimistic attitude), or an average.

\section{Related Work}
\label{sec:relatedWork}

Weighted versions of rank correlation measures have not only been studied in statistics but also in other fields, notably in information retrival \cite{Yilmaz08,Kaye1973,CostaSoares2005,Maturi2008}.
Most of them are motivated by the idea of giving a higher weight to the top-ranks: in information retrieval, important documents are supposed to appear in the top, and a swap of important documents should incur a higher penalty than a swap of unimportant ones.

Kaye \cite{Kaye1973} introduced a weighted, non-symmetric version of Spearman's rho coefficient. 
Costa and Soares \cite{CostaSoares2005} proposed a symmetric weighted version of Spearman's coefficient resembling the one of Kaye. Another approach, based on average precision and called \textit{AP correlation}, was introduced by Yilmaz \textit{et al.} \cite{Yilmaz08}.
Maturi and Abdelfattah \cite{Maturi2008} define weighted scores $W_i = w^i$ with $w \in \left(0,1\right)$
and compute the Pearson correlation coefficient on these scores. All four measures give higher weight to the top ranks.

Two more flexible measures, not restricted to monotone decreasing weights, have been proposed by Shieh \cite{Shieh1998} and Kumar and Vassilivitskii \cite{Kumar2010}.
In the approach of Shieh \cite{Shieh1998}, a weight is manually given to every occurring concordance or discordance through a symmetric weight function $w: [N] \times [N] \longrightarrow \mathbb{R}_+$:
\begin{equation}
\tau_w = \frac{\sum_{i<j}{w_{ij}C_{ij}} - \sum_{i<j}{w_{ij}D_{ij}}}{\sum_{i<j}{w_{ij}}} = \frac{\sum_{i<j}{w_{ij}(C_{ij}} - D_{ij})}{\sum_{i<j}{w_{ij}}}.
\label{eq:shieh1}
\end{equation}
The input parameter for $w$ are the ranks of a reference ranking $\pi_{ref}$, which is assumed to be the natural order $\left(1, 2, 3, \ldots, N\right)$. Therefore, this approach is not symmetric.
To handle the quadratic number of weights, Shieh proposed to define them as $w_{ij} = v_i v_j$ with $v_i$ the weight of rank $i$.

Kumar and Vassilivitskii \cite{Kumar2010} introduce a generalized version of Kendall's distance.
Originally, they proposed three different weights: element weights, position weights, and element similarities.
The three weights are defined independently of each other, and each of them can be used by its own for weighting discordant pairs.
Here, we focus on the use of position weights.
Like in our approach, Kumar and Vassilivitskii define $N-1$ weights $\delta_i \geq 0$, which are considered  as costs for swapping two elements on adjacent positions $i+1$ and $i$.
The accumulated cost of changing from position $1$ to $i \in \{ 2, \ldots , N \}$ is $p_i = \sum_{j=1}^{i-1}\delta_j$, with $p_1 = 0$. Moreover,
\begin{equation}
	\bar{p}_i(\pi_1, \pi_2) = \frac{p_{\pi_1(i)} - p_{\pi_2(i)}}{\pi_1(i) - \pi_2(i)}
\end{equation}
is the average cost of moving element $i$ from position $\pi_1(i)$ to position $\pi_2(i)$; if $\pi_1(i) = \pi_2(i)$ then $\bar{p}_i = 1$.The weighted discordance of a pair $(i,j)$ is then defined in terms of the product of the average costs for index $i$ and $j$:
\begin{equation}
	\hat{D}_{\delta}(i,j) = \begin{cases}
		\bar{p}_i(\pi_1, \pi_2)\bar{p}_j(\pi_1, \pi_2) & \text{if}\ (i,j)\ \text{is discordant}\\
		0 & \text{otherwise}
	\end{cases}\ .
	\label{eq:DHatDelta}
\end{equation}
Finally, the weighted Kendall distance $K_{\delta}$ is given by 
\begin{equation}\label{eq:kv}
	K_{\delta} = \tilde{D}_{\delta}\ = \sum_{i=1}^{N-1}{\sum_{i+1}^{N}{\hat{D}_{\delta}(i,j)}}\ .
\end{equation}
Note that (\ref{eq:kv}) is indeed a distance and not a correlation measure. To ease comparison with $\tau_{\omega}$ and $\tilde{\gamma}$, we can define 
\begin{equation}
	\hat{C}_{\delta}(i,j) = \begin{cases}
		\bar{p}_i(\pi_1, \pi_2)\bar{p}_j(\pi_1, \pi_2) & \text{if}\ (i,j)\ \text{is concordant}\\
		0 & \text{otherwise}
	\end{cases}
	\label{eq:CHatDelta}
\end{equation}
as the weighted concordance of a pair $(i,j)$, and finally another weighted version of gamma: 
$$
\tilde{\gamma}_{\delta}=\frac{\tilde{C}_{\delta} - \tilde{D}_{\delta}}{\tilde{C}_{\delta} + \tilde{D}_{\delta}} \enspace .
$$

Specific emphasis on the handling of ties is put in the recent paper \cite{plai_ca21}. The authors study the problem of rank aggregation, i.e., aggregating preferences into a consensus ranking, taking into account that swapping elements on the top might be more important than swapping elements at the bottom. To this end, they propose a position weighted rank correlation coefficient for rankings with ties, which combines the structure of Emond and Mason's measure \cite{edmo_an02} with the class of weighted Kemeny-Snell distances.  

\section{Summary and Conclusion}

We introduced a new approach to weighted rank correlation based on fuzzy order relations, as well as a concrete measure called \emph{scaled gamma}. The latter allows for specifying the importance of rank positions in a quite flexible and convenient way by means of a scaling function. Thanks to the underlying formal foundation, such a scaling function immediately translates into a concrete version of our measure, in which the rank positions are processed within an appropriate weighting scheme.  

In future work, we plan to analyze the usefulness of our new approach to rank correlation by means of experimental studies, both with synthetic and real data. More specifically, we plan to compare our measure with others on various applications and data analysis tasks, such as ranking or clustering. 
Moreover, let us again highlight that our extension of gamma is actually not a single measure but a family of measures, which is parameterized by the weight function $w$ as well as the generalized logical conjunction (t-norm) used to define concordance and discordance. While the former will typically be specified as an external parameter by the user, the (fuzzy) logical operators offer an interesting degree of freedom that could be used to optimally adapt the measure to the application at hand. Again, this is an interesting direction for future work. Finally, going beyond the gamma coefficient, we also intend to apply our generalization to other rank correlation measures.   

A Python implementation\footnote{This implementation was provided by Hoang Cong Thanh, whose contribution is kindly acknowledged by the authors.} of our weighted rank correlation measure can be found here: \url{https://github.com/KIuML/weighted_rank_correlation}

%\bibliographystyle{plain}
%\bibliography{SGrefs}

\end{document}